\documentclass[aps,prl,showpacs,twocolumn,superscriptaddress,amsmath,amssymb,floatfix]{revtex4}
\usepackage{tabularx}
\usepackage{bm}
\usepackage{epsfig,subfigure}
\usepackage{multirow}
\usepackage{graphicx}
\usepackage{color}
\usepackage{amsfonts}
\usepackage{exscale}
\usepackage{amsbsy}
 \usepackage[colorlinks,urlcolor=blue,linkcolor=blue,anchorcolor=blue,citecolor=blue]{hyperref}

\begin{document}
\title{Chiral Spin Liquid In a Frustrated Anisotropic Kagome Heisenberg Model}

\author{Yin-Chen He}
\affiliation{Department of Physics, State Key Laboratory of Surface Physics and Laboratory of Advanced Materials, Fudan University, Shanghai 200433, China}
\author{D. N. Sheng}
\affiliation{Department of Physics and Astronomy, California State University, Northridge, California 91330, USA}
\author{Yan Chen}
\affiliation{Department of Physics, State Key Laboratory of Surface Physics and Laboratory of Advanced Materials, Fudan University, Shanghai 200433, China}
\affiliation{Department of Physics and Center of Theoretical and Computational Physics, The University of Hong Kong, Pokfulam Road, Hong Kong, China}

\begin{abstract}
Kalmeyer-Laughlin (KL) chiral spin liquid (CSL) is a type of quantum spin liquid without time-reversal symmetry, and it is considered as the parent state of an exotic type of superconductor---anyon superconductor. Such exotic state has been sought for more than twenty years; however, it remains unclear whether it can exist in realistic system where time-reversal symmetry is breaking (T-breaking) spontaneously. By using the density matrix renormalization group, we show that KL CSL exists in a frustrated anisotropic kagome Heisenberg model (KHM), which has spontaneous T-breaking. We find that our model has two topological degenerate ground states, which exhibit nonvanishing scalar chirality order and are protected by finite excitation gap. Furthermore, we identify this state as KL CSL by the characteristic edge conformal field theory from the entanglement spectrum and the quasiparticles braiding statistics extracted from the modular matrix. We also study how this CSL phase evolves as the system approaches the nearest-neighbor KHM.
\end{abstract}
\maketitle

Topological order, an exotic state of matter that hosts fractionalized quasiparticles with anyonic braiding statistics,
is one of the core topics in modern condensed-matter physics \cite{Wen1990}.
Quantum spin liquid (QSL) \cite{Anderson1973} is a prominent example of topological order, which is thought to exist in some frustrated magnets \cite{Balents2010}.
Among various types of QSL \cite{Balents2010, Rokhsar1988,Senthil2000,Moessner2001, Balents2002, Read1991, Kitaev2003,Levin2003, Ran2007}, there is a class of time-reversal symmetry violating QSL called chiral spin liquid (CSL) \cite{Kalmeyer1987, Yao2007, Messio2012}.
CSL shares some similar properties with fractional quantum Hall effect, however CSL is special for its both possessing topological order and spontaneously time-reversal symmetry breaking.

The  simplest  CSL is the Kalmeyer-Laughlin (KL) CSL ($\nu=1/2$ Laughlin state) \cite{Kalmeyer1987}, in which spinons obey semionic fractional statistics. It is theoretically shown that if one dopes the KL CSL with holes \cite{Laughlin1988}, an exotic type of superconductivity---anyon superconductivity \cite{Wilczek1990}---will emerge. Inspired by the fundamental interest  and prospect of finding exotic superconductors, KL CSL has attracted  much interest \cite{Wen1989, Chen1989, haldane1995, Yang1993, Schroeter2007, Hermele2009, Bauer2013, yfwang,nielsen2012, Baskaran1989, Lee1990,Barkeshli2013, Marston1991}.
In past many years, there was no experimental or theoretical evidence supporting the existence of this state until very recently.
Several artificial models were found that can host a KL state \cite{Schroeter2007,Bauer2013, nielsen2012}. For example, one can directly induce scalar chirality order by a 3-spin parity and time-reversal-violating interaction \cite{Bauer2013} on a kagome lattice to produce KL state.
However, it remains elusive whether the KL state can exist in a system with time-reversal symmetry, which may be more closely related to real materials.
It has been suggested that KL state may exist in magnetic frustrated systems through spontaneously breaking
time-reversal symmetry~\cite{Kalmeyer1987, Wen1989}, which are  among the most difficult systems for theorists to study exactly.

In this Letter, we show that the KL state is the ground state of a frustrated anisotropic kagome Heisenberg model (KHM) by using density matrix renormalization group \cite{White1992}, a numerical method which has been proven powerful in solving quasi-one-dimensional frustrated systems  \cite{Jiang2008, Jiang2012, Yan2011, Depenbrock2012}. Compared with the previous systems with multiple spin interactions \cite{Schroeter2007,nielsen2012,Bauer2013}, the system we study here only involves two spin interactions, and the Hamiltonian has time-reversal symmetry. By the technique developed in Ref. \cite{Cincio2013,He2013}, we find two topologically degenerate ground states, both of which break time-reversal symmetry spontaneously and exhibit a nonvanishing scalar chirality order. We also get a finite energy excitation gap and small correlation length, which support that
we have  a gapped phase. Furthermore, the entanglement spectrum of the ground states fits the edge conformal field theory of the KL state. Last but not least, we calculate the modular matrix using the two ground states \cite{ Zhang2012, Cincio2013,Zaletel2013}, which gives the braiding statistics \cite{Wen1990} of emergent anyons that is the same as what is expected for the KL state. To the best of our knowledge, this is the first model that breaks time-reversal symmetry spontaneously and hosts a KL CSL.  We also show  how the system evolves as it approaches the nearest-neighbor KHM.

\emph{Model Hamiltonian.---}We study a frustrated anisotropic KHM, whose Hamiltonian is
\begin{equation}
H= J\sum_{\langle i, j\rangle} \bm S_i \cdot \bm S_j+J_2\sum_{\langle\langle i, j\rangle\rangle} S^z_i S^z_j + J_3\sum_{\langle\langle\langle i,j\rangle\rangle\rangle} S^z_i S^z_j,
\end{equation}
where the nearest-neighbor interaction $\langle i, j\rangle$ is the isotropic Heisenberg interaction, and the second $\langle\langle i, j\rangle\rangle$ and third $\langle\langle\langle i, j\rangle\rangle\rangle$ nearest-neighbor interactions are Ising-type interactions with the same magnitude, $J_2=J_3=J'$. All the interactions are antiferromagnetic, and we take $J=1$. To establish the nature of the CSL here, we mainly focus on the point $J'=1$, where we find the CSL is very robust.

We calculate the kagome lattice spin system wrapped on a cylinder with YC or XC geometry (see supplementary materials). Both finite DMRG and infinite DMRG (iDMRG) \cite{McCulloch2008} are used, and those two different algorithms give almost the same results with differences around the truncation error (energy $\sim 10^{-6}$, entropy $\sim 10^{-4}$) (see supplementary materials).
The CSL spontaneously breaks time-reversal symmetry; thus, we use both real and complex variable codes, although the Hamiltonian is real. We mainly focus on the system with even number of sites on column where the finite size effect is smaller.
We have kept up to  8000 states in the DMRG simulation, and the truncation error is smaller than $10^{-6}$ for YC8, XC8 cylinders and $5\times10^{-5}$ for YC12, XC12 cylinders. These truncation errors are small enough to obtain the
highly accurate results for the  gapped CSL we find.

\begin{figure}[h]
\centering
\includegraphics[width=0.48\textwidth]{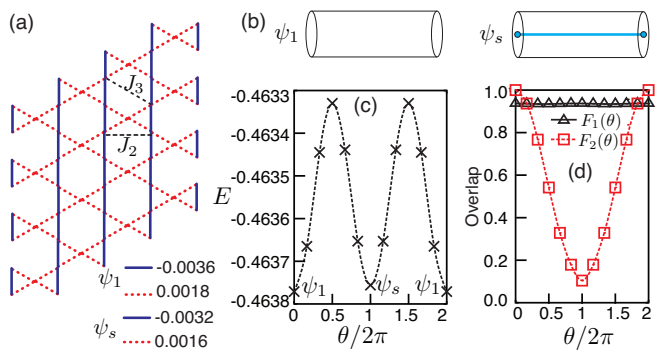} \caption{\label{fig:kagome_lattice}(Color online) (a) Illustration of YC8 ($L_y=4$) kagome lattice. Here we plot the bond spin correlation distribution of two ground states, $\langle \bm S_i \cdot \bm S_j\rangle -e_\alpha$, where $e_\alpha$ is the average of the bond spin correlations, $-0.206450$ for $\psi_1$ and $-0.206486$ for $\psi_s$. (b) Two topological degenerate ground states $\psi_1$ and $\psi_s$ in CSL. Response of YC8 cylinder under flux insertion: (c) The energy evolution. (d) The one column overlap between states at different fluxes $F_1(\theta)=|\langle\psi(\theta)| \psi(\theta+\pi/3)\rangle|$ and $F_2(\theta)=|\langle\psi(0)| \psi(\theta)\rangle|$. Here, $\psi(0)=\psi_1$ and $\psi(2\pi)=\psi_s$.}
\end{figure}

\begin{figure}[h]
\centering
\includegraphics[width=0.4\textwidth]{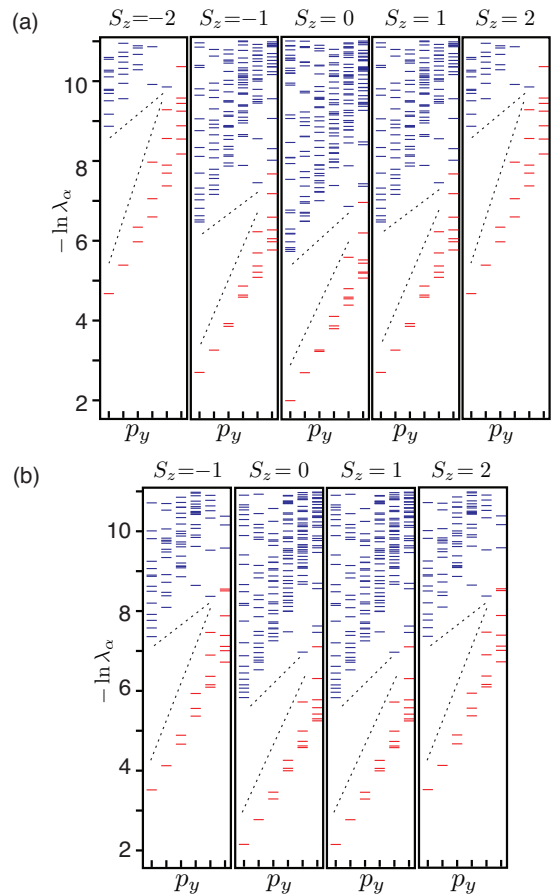}
 \caption{(Color online) \label{fig:ES} Entanglement spectrum of YC12 cylinder ($L_y=6$) (a) $\psi_1$ and (b) $\psi_s$. The horizontal axis is the momentum along the $y$ direction, $p_y=0, 2\pi/6, \cdots, 5\times2\pi/6$ (up to a global shift). The leading spectra are marked in red, above which there is a spectrum gap marked with a dashed line. Other geometry and size samples have similar results.}
\end{figure}

\emph{Degenerate ground states and their properties.---}Topological degeneracy is usually defined on a torus, but for an infinite cylinder (or sufficiently long
cylinder), one can also have a full set of topological degenerate states in the bulk of the system.  Using the technique developed in Ref. \cite{He2013}, we can get two topological degenerate ground states--$\psi_1$ and $\psi_s$--which are distinguished by the absence or presence of a spinon line, as shown in Fig. \ref{fig:kagome_lattice}(b). We obtain $\psi_1$ naturally in the conventional DMRG simulation, and get $\psi_s$ by creating edge spinons (pin or remove one site). On the other hand, by inserting a $2\pi$ flux in the cylinder, we find that $\psi_1$ ($\psi_s$) adiabatically evolves into $\psi_s$ ($\psi_1$) (Fig. \ref{fig:kagome_lattice}(c)), which indicates that the $2\pi$ flux insertion will pump a spinon from one edge to the other edge, a property for the $\nu=1/2$ fractional quantum Hall state. We also calculate the one column overlap \cite{He2013} between states at different fluxes, as shown in Fig. \ref{fig:kagome_lattice}(d). From the overlaps $F_1(\theta)=|\langle\psi(\theta)| \psi(\theta+\pi/3)\rangle|$, we know the state evolves adiabatically as the flux is inserted. Nevertheless, $F_2(\theta)=|\langle\psi(0)| \psi(\theta)\rangle|$ decreases as the inserted flux $\theta$ increases from $0$ to $2\pi$. In particular, $\psi_1$ and $\psi_s$ have a very small overlap ($f\sim 0.1$), supporting the idea that they are distinct states. The one column overlap ($f$) between two degenerate states can also give us the correlation length of spinon $\varepsilon_s=-1/\log f\sim 0.4$ \cite{He2013}. From the symmetry properties of the entanglement spectrum \cite{He2013} (also see Fig. \ref{fig:ES}), we also find that $\psi_1$ and $\psi_s$ are different topological degenerate ground states related to the absence or presence of the spinon line. The energy and entropy of the two states are very close (Table \ref{table}), and their difference drops quickly as the system size increases. It is consistent with the theory of topological degeneracy, where the energy difference is expected to vanish exponentially with system width, $\Delta E\sim\exp(-L_y/\varepsilon_s)$.
\begin{table}[h]
\caption{\label{table}The energy $E$, entropy $S$, correlation length $\xi$, chirality order $\chi_\vartriangleright$, $\chi_\vartriangleleft$ , singlet gap $\Delta_s$, triplet gap $\Delta_t$. The energy and entropy of YC12 as well as XC12 cylinders have been extrapolated versus truncation error.}
 \centering
 \begin{tabular}{ccccccccc}
  \hline
  \hline
 State & $E$ & $S$ & $\xi$ & $\chi^\vartriangleright$  & $\chi^\vartriangleleft$ & $\Delta_s$ & $\Delta_t$  \\
  \hline
  $\psi_1$, YC8 & -0.463771 & 2.880 & 1.08  & 0.09 & 0.09 & 0.17 & 0.40 \\
  $\psi_s$, YC8 & -0.463756 & 2.875 & 0.85  & 0.09 & 0.09 & 0.16 & 0.40 \\
 $\psi_1$, YC12 &-0.46356936& 4.3688 & 0.60 & 0.104 & 0.104  & 0.24 & 0.42\\
$\psi_s$, YC12 &-0.46356941& 4.3687 & 0.60 & 0.104 & 0.104  & 0.24 & 0.42 \\

  $\psi_1$, XC8 &  -0.463630  & 2.711   & 0.79   & 0.095  & 0.095  & 0.14  & 0.35  \\
  $\psi_s$, XC8 &  -0.463643 & 2.719  & 0.87   & 0.093 & 0.093 & 0.15  & 0.35 \\
 $\psi_1$, XC12 & -0.4635381 & 4.153  &  0.55  &  0.103 & 0.103  & 0.22  & 0.41 \\
$\psi_s$, XC12 &  -0.4635382 & 4.154  & 0.55  &  0.103 & 0.103  & 0.22  & 0.41 \\
  \hline
  \hline
 \end{tabular}
\end{table}

\begin{figure}[h]
\centering
\includegraphics[width=0.48\textwidth]{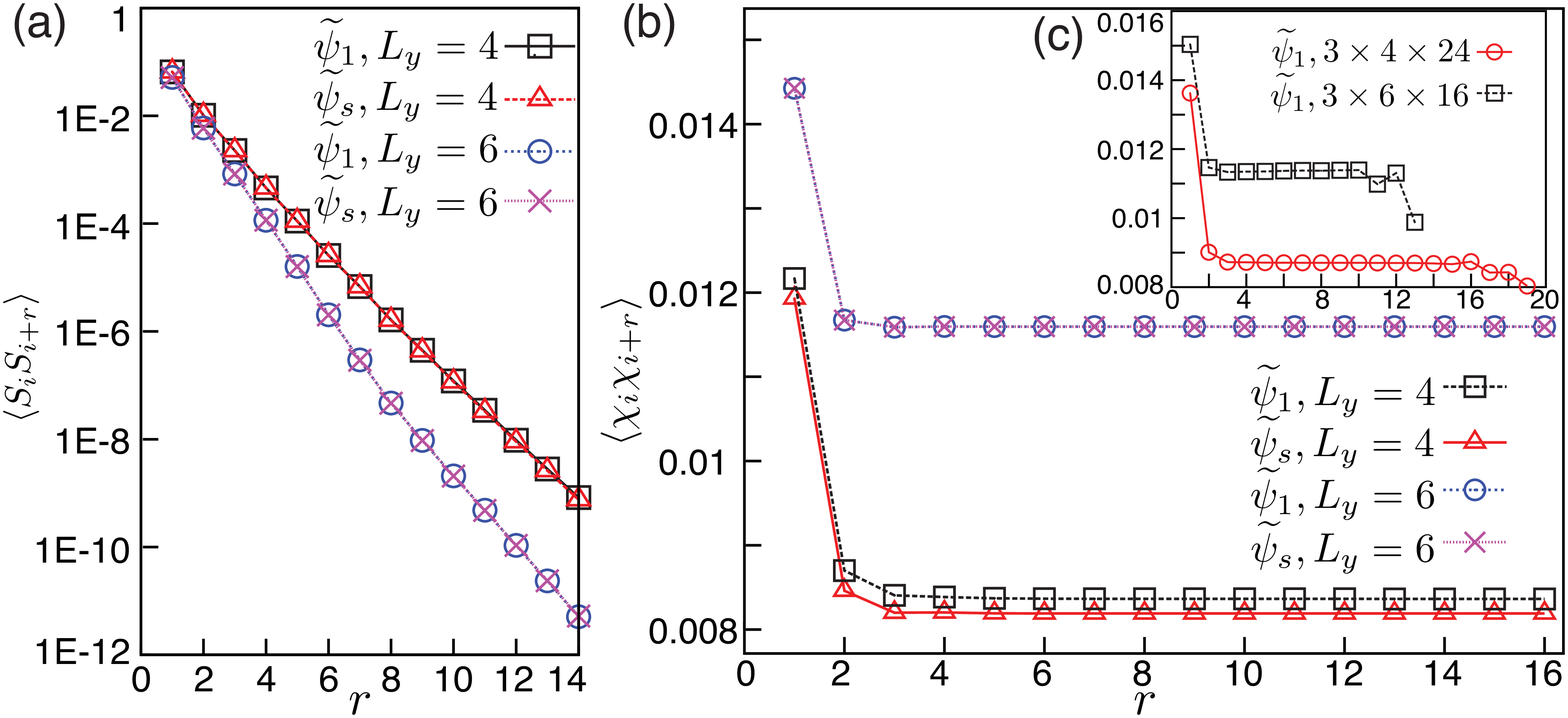} \caption{\label{fig:chiral_correlation}(Color online) (a) Spin correlation $\langle \bm S_i \cdot \bm S_{i+r}\rangle$ versus distance $r$. The chiral correlation $\langle \chi_i \chi_{i+r}\rangle$ versus the distance $r$: (b) Results from the iDMRG. (c) Results from the finite DMRG. The system sizes we calculated are $3\times 4 \times 24$ and $3\times 6 \times 16$ YC cylinders. Here $J'=1$. }
\end{figure}

To check whether the state spontaneously breaks time-reversal symmetry, we measure the scalar chirality order \cite{Wen1989}:
\begin{equation}
\chi_i^{\vartriangleright(\vartriangleleft)}=\langle\bm S_{i_1} \cdot (\bm S_{i_2} \times \bm S_{i_3}) \rangle, \quad\quad i_1,i_2,i_3 \in \vartriangleright,\vartriangleleft. \label{eq:chiral}
\end{equation}
We find $\chi$ is homogeneous on kagome lattice, where the up and down  triangles have the same chirality $\chi^{\vartriangleright}\approx\chi^\vartriangleleft$.
We also calculate the overlap between $\psi_i$ and its conjugate $\psi_i^*$, where a very small value ($3\times 4 \times 16$ YC8 cylinder $\sim 10^{-6}$, $3\times 4 \times 24$ YC8 cylinder $\sim 10^{-8}$) is obtained. The orthogonality between $\psi_i$ and $\psi_i^*$
indicates that we have two orthogonal states with opposite chirality.
We further compare these two states with the results of real variable code simulation,  we find $\psi_i$ and $\psi_i^*$  are  minimal entangled states (the superposition states $(\psi\pm \psi_i^*)/\sqrt 2$ will be  maximal entangled states) of a spontaneous time-reversal symmetry breaking system. The triplet gap and singlet gap are calculated on a finite cylinder embedded in the middle of the infinite cylinder (see supplementary materials). From the correlation length, singlet and triplet gaps, we infer the existence of a large gap between the ground state and the excited states. To confirm the absence of magnetic order, we plot the spin correlation $\langle \bm S_i \cdot \bm S_{i+r}\rangle$ in Fig. \ref{fig:chiral_correlation}(a) which clearly exhibits an exponentially decaying behavior. The nearest bond spin correlation is very homogeneous with a difference of around $1\%$ (Fig. \ref{fig:kagome_lattice}(a)), so we can exclude the translational symmetry breaking phase, such as a valence bond crystal. All these properties suggest a CSL phase.

As a self-consistency check, we also calculate properties of the system using a real-variable code. For a time-reversal symmetry breaking phase, a real-variable code will yield a maximal entangled state $\widetilde\psi\sim (\psi\pm \psi^*)/\sqrt 2$.
$\widetilde \psi$ has a doubly degenerate entanglement spectrum, and its entropy is $\ln 2$ larger than that of $\psi$. For the time-reversal symmetric state, $\widetilde\psi$, the scalar chirality order in Eq. \ref{eq:chiral} is $0$, but we can extract it from the chiral correlation function $\langle \chi_{i} \chi_{i+r}\rangle=\langle[\bm S_{i_1} \cdot (\bm S_{i_2} \times \bm S_{i_3})][ \bm S_{{(i+r)}_1} \cdot (\bm S_{{(i+r)}_2} \times \bm S_{{(i+r)}_3})] \rangle$. Clearly observed in Fig. \ref{fig:chiral_correlation}(b) is a long-range chiral correlation, $\sqrt{\lim_{r\rightarrow \infty} \langle \chi_i \chi_{i+r}\rangle }\approx 0.09$ (YC8) and $0.1$ (YC12).

\emph{Entanglement spectrum.---}The entanglement spectrum can serve as a fingerprint for a topological chiral phase \cite{Li2008}. In the DMRG simulation, one can naturally obtain the entanglement spectrum along a vertical cut. To see whether the entanglement spectrum fits the conformal field theory, one needs to calculate the momentum (along the $y$ direction) of each entanglement spectrum \cite{Cincio2013}. Since the system has a $U(1)$ symmetry, each spectrum has an $S_z$ quantum number. From Fig. \ref{fig:ES}, one can see the  pattern of $\psi_1$ is symmetric about
the positive and negative $S_z$, whereas that of $\psi_s$ is symmetric about  $S_z=0$ and $S_z=1$. This distinct symmetry pattern is the consequence of the absence or presence of the spinon line as shown in Fig. \ref{fig:kagome_lattice}(b). Moreover, one can clearly observe an entanglement spectrum gap, and the degeneracy pattern ($\{1,1, 2, 3, 5, \cdots\}$) of the low-lying spectra. This is in agreement with the edge conformal field theory of the KL state \cite{Cincio2013}.

\emph{Braiding statistics from modular matrix.---}Next, we use $\psi_1$ and $\psi_s$ to calculate the modular matrix \cite{Wen1990}, $\mathcal S, \mathcal U$, which contains full information of a topological ordered phase. For a topological chiral state, the modular matrix is
\begin{equation}
\mathcal S=\left( \begin{matrix} S_{11} & S_{1s} \\ S_{s1} & S_{ss}\end{matrix}\right), \quad\quad \mathcal U=e^{-i (2\pi/24)c} \left( \begin{matrix} h_1 & 0 \\ 0 & h_s \end{matrix} \right).
\end{equation}
$S_{1i}=S_{i1}=d_i/D$, where $d_i$ is the quantum dimension of quasiparticle (type $i$), and $D=\sqrt{\sum_i d_i^2}$ is the total quantum dimension. The topological entanglement entropy \cite{Kitaev2006, Levin2006} can also give the quantum dimension \cite{footnote0}.
The entry $S_{ij}$ will give the braiding statistics of the anyon model. Furthermore, $c$ is the central charge of the system, $h_i$ is called the topological spin which determines the self-statistics of the type-$i$ anyon. For Abelian anyons, one has $d_i=1$ and $S_{ij}=\exp (i \theta_{ij} )/\sqrt D$ where $\theta_{ij}$ is the phase coming from a type-$i$ anyon encircling a type-$j$ anyon.

Following the procedure outlined in Ref. \cite{Zhang2012, Cincio2013}, we get the modular matrix of YC8 cylinder using Monte Carlo sampling \cite{Sandvik2007},
\begin{align}
\mathcal S&=\frac{1}{\sqrt 2}\left( \begin{matrix} 1 & 1 \\ 1 & -1 \end{matrix} \right)+\frac{10^{-2}}{\sqrt 2}\left( \begin{matrix} -0.42 & -2.2 \\ -1.26 & 0.76-0.15i \end{matrix} \right)
\end{align}
and
\begin{align}
\mathcal U&=e^{-i (2\pi/24)} \left( \begin{matrix} 1 & 0 \\ 0 & i \end{matrix} \right)\times  \left( \begin{matrix} e^{0.011i} & 0 \\ 0 & e^{-0.006i}  \end{matrix} \right)
\end{align}
Generally, from the modular matrices, one can know: (1) there are two types of quasiparticles, a trivial vacuum $1$ and a spinon $s$---they are all Abelian quasiparticles with quantum dimension $d_i=1$; (2) the spinon has semionic braiding statistics relative to itself; (3) the fusion rules are $1\times 1=s\times s=1$, $1\times s=s\times 1=s$; (3) the topological spin of the two quasiparticles are $h_1=1$, $h_s=i$; and (4) the central charge is $c=1$. From these results, we can conclude this CSL state is KL state.

It is interesting to know the mechanism for the KL CSL phase. An intriguing proposal for the emergence of CSL on kagome lattice is given in Ref. \cite{Messio2012}. They found that a $J_1\sim J_2 \sim J_3$ SU(2) Heisenberg interaction on kagome lattice will lead to a classical chiral state with 12 spins pointing toward the corners of a cuboctahedron. Quantum fluctuation will disrupt the classical order and result in a CSL state. However, our Ising anti-ferromagnetic interaction does not favor such a complex classical state, as it only prefers to align spins anti-parallel along the $z$-direction. Moreover, the CSL proposed in Ref. \cite{Messio2012} is a Z$_2$ CSL, different from the KL-CSL we find here. We think the CSL here comes from quantum frustration on the kagome lattice as originally proposed by KL \cite{Kalmeyer1987}, and the Ising anti-ferromagnetic interactions only play the role of stabilizing the CSL state
 through enhancing the repulsive interactions between spin particles. In a parallel work,  CSL has also been found in an SU(2) rotational invariant Heisenberg model~\cite{ssgong13} on kagome lattice.
 It is interesting to study whether real material relevant interactions \cite{Jeschke2013} may also stabilize the KL state on the kagome lattice, which will help to find KL CSL or even anyon superconductor in real materials like Herbertsmithite ZnCu$_3$(OH)$_6$Cl$_2$.

\emph{Chiral spin liquid and nearest neighbor KHM.---}In the following, we study how the CSL evolves into the limit ($J'=0$), nearest-neighbor KHM,
where a possible $Z_2$ SL was previously discovered~\cite{Yan2011}. To systematically study the state evolution, we use both real and complex variable codes, where we find the former has a bias towards a time-reversal invariant spin liquid (TSL) \cite{footnote1}, whereas the latter has a bias towards a CSL (see supplementary materials). In the whole region $J'\in [0,1.2]$, the two degenerate ground states---$\psi_1$ and $\psi_s$---always exist,
moreover, due to the finite size effect, these two states behave differently as $J'$ decreases.

\begin{figure}[h]
\includegraphics[width=0.48\textwidth]{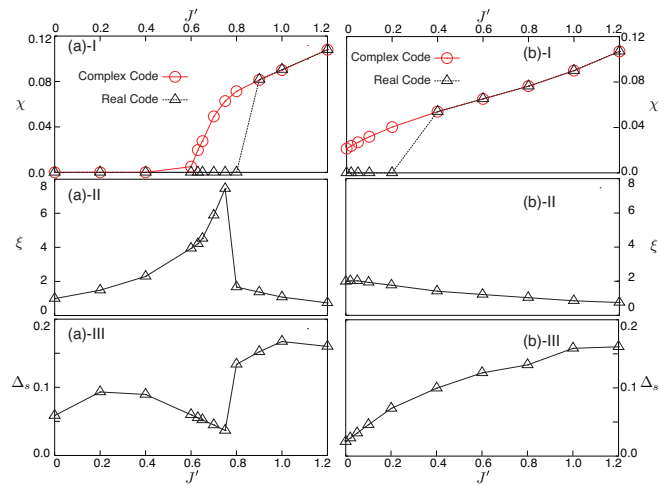} \caption{\label{fig:evolution} (Color online) $J'$-dependent behavior of the two topological degenerate sectors (a) normal sector, $\psi_1$ (b) spinon sector, $\psi_s$. Here, we show results of calculations for the YC8 cylinder: (I) the chirality order, $\chi$ (Eq. \ref{eq:chiral}), where for the real code, we define $\chi=\sqrt{\lim_{r\rightarrow \infty} \langle \chi_i \chi_{i+r}\rangle }$; (II) the correlation length $\xi$ of the ground state; (III) the singlet gap $\Delta_s$ of the ground state.}
\end{figure}

Because of the bias of the real and complex code, we find in some regions (transition region), two codes give different states: TSL and CSL. As shown by the scalar chirality order, the transition region for $\psi_0$ is around $(0.6,0.8)$, and that for $\psi_s$ is around $(0.0,0.4)$.
 The energy difference of TSL and CSL is very small ($\sim 10^{-5}$), indicating the strong competition between these two states.
We find that, for the $\psi_0$, TSL has a lower energy than CSL when $J'\lesssim 0.7$; in contrast, for $\psi_s$, the CSL always has a slightly
lower energy than TSL. The CSL in the spinon sector may be related to the excited state with non-zero Chern number on nearest neighbor KHM found in the exact diagonalization study \cite{Waldtmann1998}.
The two sectors may also behave similarly as the system size increases. The quantum phase transition between the two topological phases, CSL and TSL, may result in new intriguing critical behavior, which may explain the results we obtain in the transition region.  We leave these open and challenging topics to the future study.

\emph{Conclusion.---}Using the DMRG, we numerically show that KL CSL is the ground state of a frustrated anisotropic KHM. We provide various evidences to support KL CSL's existence, especially the presence of finite scalar chirality order, topological degeneracy, large energy gap and the verification of semionic fractional statistics of the spinons. Our work  shows that KL CSL can emerge in a realistic model for the first time
which only involves time-reversal invariant two-spin exchange interaction terms. Furthermore, we study how the system evolves into the nearest neighbor KHM, and provide some explanation of the competing evidence between CSL and $Z_2$ SL.
It would be interesting to  study much larger systems using different approaches
(like tensor network or projected wavefunction)  to  further 
 establish the topological phase and exclude the  possibility of a weak order.

\textit{Acknowledgments.---} D.N.S. thanks Shoushu Gong and Wei Zhu for extensive discussions.  This work was supported by the State Key Programs of China (Grant No. 2012CB921604 and 2009CB929204), the National Natural Science Foundation of China (Grant No. 11074043 and 11274069) (YCH and YC), and the US National Science Foundation under grant DMR-0906816 (DNS).

\newpage

\onecolumngrid

\begin{center}
{\bf \large Supplementary Material of ``Chiral Spin Liquid In a Frustrated Anisotropic Kagome Heisenberg Model"}
\end{center}

\section{Cylinder Geometry of kagome lattice}
\begin{figure}[h]
\centering
\includegraphics[width=0.5\textwidth]{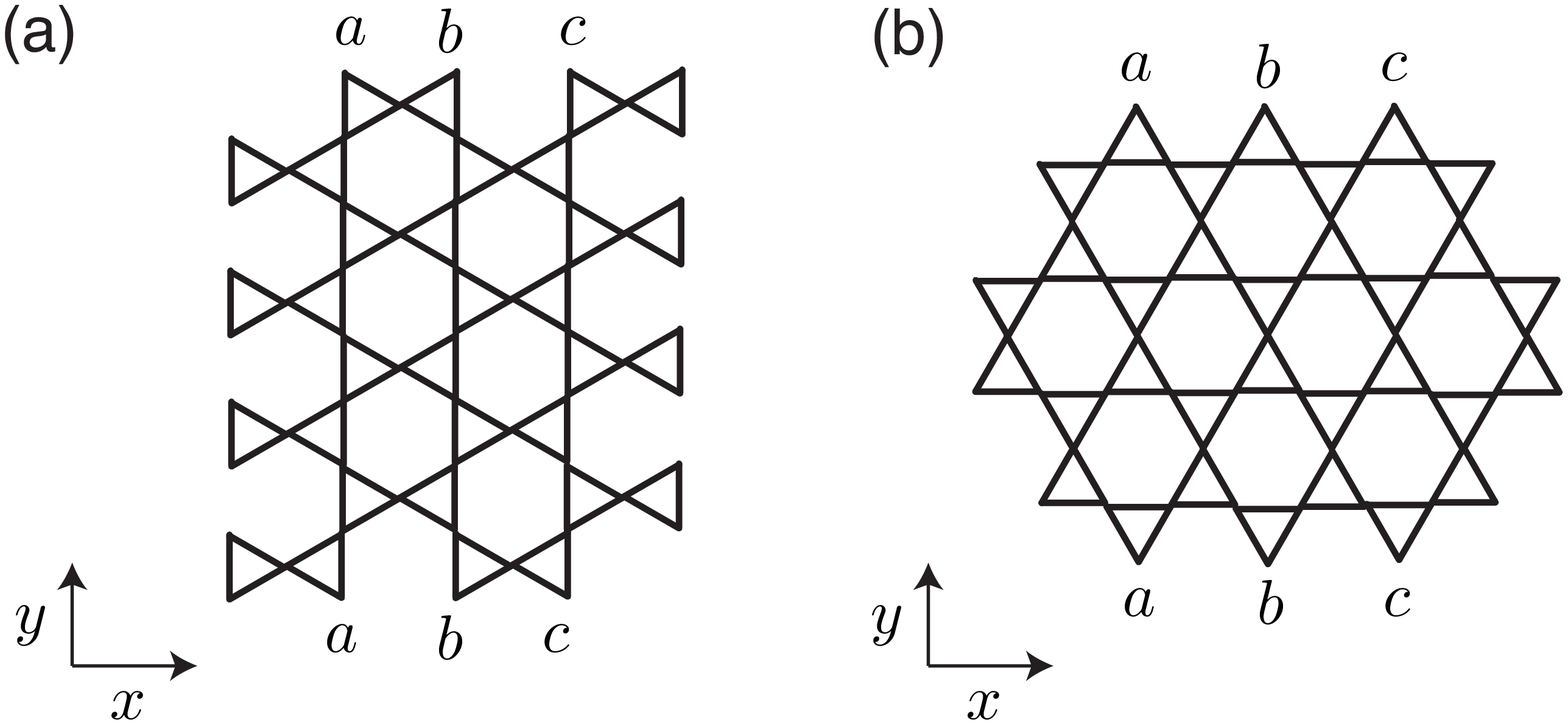} \caption{\label{fig:geo_sup} (a) YC8 cylinder. (b) XC8 cylinder.}
\end{figure}

We have two ways to wrap kagome lattice on the cylinder, which we call YC and XC cylinders as shown in Fig. \ref{fig:geo_sup}. We choose open boundary condition along $x$ direction, and periodic boundary condition along $y$ direction (with identical sites labeled by the same alphabet).

\section{Comparison between iDMRG and finite DMRG}
In the finite DMRG simulation, one first runs the iDMRG algorithm to reach the required system size, and then the finite DMRG sweep is used to improve accuracy. This is a widely used routine procedure. Usually, it is thought that using the iDMRG alone without the finite DMRG sweep will give poor accuracy, although this is not the case for the iDMRG algorithm. In the finite DMRG simulation, the sweep is necessary because the system's length is too small. In the iDMRG simulation, one has a cylinder or chain growing to a very large length until convergence is reached. The iDRMG gives the fixed-point wave function of the center of the chain or cylinder, which is exactly the same as that obtained in the finite DMRG simulation for a gapped phase. Moreover, if one is only interested in the fixed-point wave function of the center in a long chain or cylinder, the finite sweep is not efficient, because the accuracy of the wave function far away from the center is not important for a gapped phase (the perturbation does not propagate far owing to the small correlation length).

In the following, we give a detailed comparison between the iDMRG and finite DMRG. We set a very high accuracy bar here: for each $m$ state kept, we demand that the total energy and entropy difference between the $n_{th}$ and $(n+1)_{th}$ sweeps (iterations) satisfy $|E_n-E_{n+1}|/E_{n+1}<10^{-7}$ and $|S_n-S_{n+1}|/S_{n+1}<10^{-5}$. The performance and results are shown in Table \ref{table:comparison_sup}. We use the finite DMRG to calculate the $3\times 4 \times 16$ and $3\times 4\times 24$ cylinders. The energy is computed in the middle $3\times4\times2=24$ sites, and entropy along the middle vertical cut of the cylinder. The computational cost is defined by how many sites need to be optimized. For example, by keeping $m=1000$ states, a $3\times 4 \times 16$ cylinder needs $4$ sweeps to reach the required accuracy, so the computational cost is $192\times2\times4=1536$; for the iDMRG method, it takes $14$ iterations, hence the cost is $3\times4\times2\times14=336$. We find iDMRG not only yields a much lower computational cost, but also gives results with a little higher accuracy.
\begin{table}[h]
\caption{Comparison between finite DMRG and iDMRG. Here $J'=1$, $\varepsilon$ is the truncation error.}
 \centering
 \begin{tabular}{ccccccccccccc}
  \hline
  \hline

& \multicolumn{3}{c}{$m=1000$, $\varepsilon=3\times 10^{-5}$} & \multicolumn{3}{c}{$m=2000$, $\varepsilon=1\times 10^{-5}$} & \multicolumn{3}{c}{$m=3000$, $\varepsilon=5\times 10^{-6}$}  \\
& $E$ & $S$ & cost & $E$ & $S$ & cost & $E$ & $S$ & cost \\
$3\times4\times16$
& -0.4631511  & 2.7229  & 1536 & -0.4635612 & 2.8036 & 1920 & -0.4636618 & 2.8343 & 1536 \\
$3\times4\times24$
& -0.4631607& 2.7226 & 2304 & -0.4635719 & 2.8034 & 2880 & -0.4636741 &2.8341 &2304\\
iDMRG
 &  -0.4631632 & 2.7223 & 336 & -0.4635747 & 2.8033 & 240 & -0.4636754  & 2.8340 & 288 \\
  \hline
  \hline
 \end{tabular}
  \label{table:comparison_sup}
 \end{table}

\section{Triplet and Singlet Gap}

We use an algorithm that combines iDMRG and finite DMRG to calculate the triplet as well as singlet gap. Firstly, we use iDMRG to obtain converged wave-function of an infinite cylinder, then we cut the infinite cylinder into to halves, insert a $3\times L_y \times L_y$ cylinder in the system. The left ($L$) and right ($R$) semi-infinite cylinder can be considered as environment (boundary conditions), and we calculate the energy of ground state  $E_0(S_z=0)$, spin-1 sector $E_0(S_z=1)$ and singlet excited-sector $E_1(S_z=0)$ within the inserted $3\times L_y \times L_y$ cylinder. Finally, we obtain the triplet gap $\Delta_t=E_0(S_z=1)-E_0(S_z=0)$ and singlet gap $\Delta_s=E_1(S_z=0)-E_0(S_z=0)$. 

\begin{figure}[h]
\centering
\includegraphics[width=0.4\textwidth]{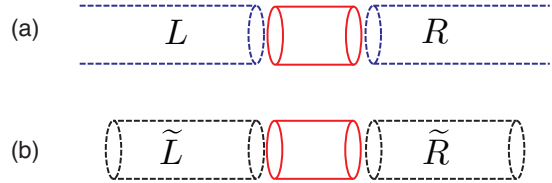} \caption{\label{fig:gap_ill_sup} (a) iDMRG--finite DMRG combined algorithm (b) Finite DMRG algorithm.}
\end{figure}

This algorithm is similar as the one used in finite DMRG, where one obtains the triplet and singlet gap with sweeping in the middle of a finite cylinder to minimize the boundary effect. The only difference is that the boundary environment we use come from the iDMRG simulation, while finite DMRG uses environment from finite cylinder simulation.

\section{Chiral Spin Liquid and Time-Reversal Invariant Spin Liquid}
\begin{figure}[h]
\centering
\includegraphics[width=0.4\textwidth]{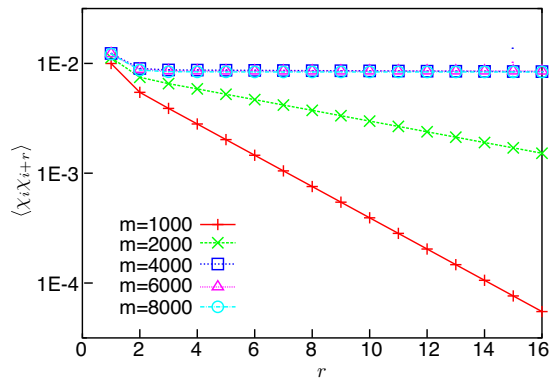} \caption{\label{fig:chiral_cor_sup} Chiral correlation $\langle \chi_i \chi_{i+r}\rangle$ versus the distance $r$. Here we use iDMRG, with $J'=1$, $L_y=4$ and the state is in the normal sector ($\psi_1$).}
\end{figure}

The real-variable code has a bias towards a time-reversal invariant (T-invariant) state, which can be observed in Fig. \ref{fig:chiral_cor_sup}. With a small number of states kept, the real-variable code yields a state with short-range chiral correlation; while one keeps a larger number of states, the state develops a long-range chiral correlation.

\begin{thebibliography}{99}
\bibitem{footnote0} Due to the difficulty of extrapolating the converged entanglement  entropy for larger systems
(YC12, XC12) and the finite size effect, the topological entanglement entropy (TEE) extrapolated from our data has a deviation from the theoretical value $\log \sqrt 2$. However,  a near accurate TEE term is obtained for the CSL in SU2 model\cite{ssgong13} by keeping
around  16000 states.
\bibitem{footnote1} A state from a real-variable code is always time-reversal symmetric. The time-reversal invariant state here means a state with an exponentially decaying chiral correlation instead of a long-range chiral correlation in Fig. \ref{fig:chiral_correlation}.
\end{thebibliography}
\end{document}